\documentclass[aip,numerical,graphicx]{revtex4-1}
\usepackage{graphicx}% Include figure files

\draft % marks overfull lines with a black rule on the right

\def\nabp{\nabla_{\perp}}
\newcommand {\be}{\begin{equation}} % start equation
\newcommand{\ee}{\end{equation}}    % end equation

\begin{document}

%\preprint{}

\title{Nonlinear three wave interaction in pair plasmas} %Title of paper



\author{J. Vranjes} %and S. Poedts}
\email[]{Jovo.Vranjes@wis.kuleuven.be}%; Stefaan.Poedts@wis.kuleuven.be}
\author{S. Poedts}
\email[]{Stefaan.Poedts@wis.kuleuven.be}
%\thanks{}
\affiliation{Centre for Plasma Astrophysics, and Leuven Mathematical Modeling and Computational Science Research Centre
 (LMCC),  Celestijnenlaan 200B, 3001 Leuven,  Belgium.}


\date{\today}

\begin{abstract}
It is shown that nonlinear  three-wave interaction, described by vector-product type nonlinearities, in pair plasmas
implies much more restrictive conditions for a double energy transfer, as compared to electron-ion plasmas.
\end{abstract}

\pacs{52.35.Mw, 52.30.Ex, 52.27.Ep  }% insert suggested PACS numbers in braces on next line

\maketitle %\maketitle must follow title, authors, abstract and \pacs

%\section{Introduction}
%\label{}
%\section{Introduction}
An electron-ion plasma, with a density gradient perpendicular to the ambient magnetic field vector, supports drift
waves driven by the density gradient. In the nonlinear regime, the vector-product-type nonlinearity leads to a
three-wave interaction which allows for both a direct and an inverse energy transfer \cite{hm, km, w}, i.e., the
transport of energy towards both shorter and longer wavelengths. A similar behavior is obtained also in rotating
self-gravitating astrophysical clouds \cite{v1}, where the inverse energy transfer implies the formation of large-scale
structures on time scales much shorter than the gravitational contraction.

In the case of pair plasmas (electron-positron\cite{z, iw} or pair ion plasmas), instead of the drift mode one obtains
convective cells \cite{v2}.  In the past, the physics of electron-positron plasmas  has been investigated mainly
related to problems of active galactic nuclei and neutron stars. Nevertheless, experimental techniques have been
developed  for collecting and keeping positrons, and thus the anti-matter plasma has been the subject of experimental
investigations as well. However, the problem of particle annihilation, that is inherent in electron-positron plasmas,
is absent in the recently experimentally produced pair-ion plasmas \cite{h1, h2, h3, h4}. This experimental success has
triggered an increased activity in the field in the past a few years \cite{v3, k1, v2a, s4, v5,gil}.

In the present work, we shall study the nonlinear wave interaction in pair plasmas. Hence, we assume a plasma with two
components of equal mass and opposite charge (electron-positron or pair-ion), and  we use the continuity and momentum
equations for any of the two species, in the form
\be
\frac{\partial}{\partial t} \left(\frac{n_1}{n_0}\right) + \nabla_\bot \cdot \vec v_{\bot 1} + \frac{\partial
v_{z1}}{\partial z} + \vec v_{\bot 1}\cdot \nabla_\bot n_0 + \frac{n_1}{n_0} \nabla\cdot \vec v_1 + \vec v_1
\frac{\nabla n_1}{n_0}=0, \label{e1} \ee
\be
\left(\frac{\partial}{\partial t }+ \vec v_1\cdot \nabla\right)\vec v_1=\frac{q}{m} \left(-\nabla \phi_1 + \vec
v_1\times \vec B_0\right). \label{e2} \ee
The indices  $0$ and $1$ here denote  equilibrium and perturbed  quantities, respectively. We study low frequency
perturbations $\sim \exp(-i \omega t + i \vec k_\bot \vec r + i k_z z)$, $|\partial/\partial t|\ll |\Omega|=|q|B_0/m$,
propagating nearly perpendicularly with respect to the magnetic field vector $\vec B_0=B_0 \vec e_z$, i.e.,
$|k_\bot|\gg |k_z|$, and we allow for the presence of the equilibrium perpendicular density gradient $\nabla_\bot n_0$.
It will be shown that the density gradient effects play no role in the present problem.

From Eq.~(\ref{e2}) we obtain approximately \be
\vec v_{\bot 1}= \frac{1}{B_0} \vec e_z\times \nabla_\bot \phi_1 - \frac{1}{B_0 \Omega} \left(\frac{\partial}{\partial
t} + \frac{1}{B_0}\vec e_z \times \nabla_\bot \phi_1\cdot \nabla_\bot\right) \nabla_\bot \phi_1, \label{e3} \ee \be
\left(\frac{\partial}{\partial t} + \frac{1}{B_0}\vec e_z \times \nabla_\bot \phi_1\cdot \nabla_\bot\right) v_z=
-\frac{q}{m} \frac{\partial \phi_1}{\partial z}. \label{e4} \ee
Using these two in Eq.~(\ref{e1}), for any of the two species one may write
\[
 {\cal L} n_1 =\frac{n_0}{B_0 \Omega}{\cal L} \nabla_\bot^2 \phi_1 - n_0 \frac{\partial v_{z1}}{\partial z} -
\frac{n_0}{B_0}(\vec e_z \times \nabla_\bot \phi_1) \cdot \nabla_\bot \log n_0
\]
\be -n_1 \nabla_\bot\cdot \vec v_{\bot 1}
 -
n_1 \frac{\partial v_{z1}}{\partial z} - v_{z1} \frac{\partial n_1}{\partial z}. \label{e5}
\ee
Here,
\[
{\cal L}=\frac{\partial}{\partial t} + \frac{1}{B_0}\vec e_z \times \nabla_\bot \phi_1\cdot \nabla_\bot.
\]
Now, writing two equations (\ref{e5}) for the two species and equating the corresponding expressions in view of the
assumed quasi-neutrality, it is seen that the terms with the equilibrium density gradients exactly cancel each other
out. The resulting combined equation for the species $a, b$ consequently reads:
 \be
 \left(\frac{1}{B_0\Omega_a}-\frac{1}{B_0\Omega_b}\right){\cal L} \nabla_\bot^2 \phi_1+ \frac{\partial}{\partial z}
 (v_{bz1}-v_{az1}) =0. \label{e6}
 \ee
Applying the operator ${\cal L}$ onto Eq.~(\ref{e6}) once more, and after using Eq.~(\ref{e4}),  one obtains
 \be
\left(\frac{\partial}{\partial t} + \frac{1}{B_0}\vec e_z \times \nabla_\bot \phi_1\cdot
\nabla_\bot\right)\left[\left(\frac{\partial}{\partial t} + \frac{1}{B_0}\vec e_z \times \nabla_\bot \phi_1\cdot
\nabla_\bot\right) \nabp^2 \phi_1\right] = - \Omega^2 \frac{\partial^2 \phi_1}{\partial z^2}. \label{e7} \ee
Linearizing Eq.~(\ref{e7}) one then obtains the dispersion equation for the electrostatic convective cells as normal
modes in this pair plasma, propagating almost perpendicularly to the magnetic field vector:
\be
\omega^2=\Omega^2 \frac{k_z^2}{k_\bot^2}, \qquad \Omega=|q|B_0/m. \label{e8} \ee
 Note that using the Poisson equation instead of the quasi-neutrality will only modify the constant on the
right-hand side of Eqs. (\ref{e7}, \ref{e8}), $\Omega^2\rightarrow \Omega^2/[1+ \Omega^2/(2 \omega_p^2)]$, $\omega_p^2=
q^2 n_0/(\varepsilon_0 m)$. This modification is not essential for our work and will be neglected  implying that
$\Omega^2\ll 2\omega_p^2$.

%It can be shown that in the strongly nonlinear domain, and looking for quasi-static propagating solutions,
%Eq.~(\ref{e7}) can be solved yielding solutions in the shape of a double vortex traveling in the perpendicular and
%parallel directions by some constant velocities $u_{\bot}, u_z$.

In order to study the three-wave interaction, we use the nonlinear Eq.~(\ref{e7})
 assuming the perturbed potential (after omitting  the previously used index 1) in the form
\be
\phi(t)=\sum_{j=1}^{j=3} \left[\Phi_j(t)\exp\left(-i \omega_j t + i \vec k_j\vec r\right) + \Phi_j^*(t)\exp\left(i
\omega_j t - i \vec k_j\vec r\right)\right].\label{s} \ee
Here, $*$ denotes the complex-conjugate quantity, and we shall use also
\[
\left.\frac{\partial}{\partial t}\right|_j\rightarrow - i \omega_j + \frac{\delta }{\delta t}, \qquad |i \omega_j|\gg
|\delta /\delta t|,
\]
where $\delta/\delta t$ is  the time derivative on a   slow time-scale, as a consequence of the nonlinear three-wave
interaction. In this case, using Eq.~(\ref{e8}), from  Eq.~(\ref{e6}) one  obtains  the following expression for  time
variation of the $j$th-amplitude:
\[
\frac{\delta \Phi_j(t)}{\delta t}=\frac{i}{2 \omega_j k_j^2}\left\{\frac{1}{B_0} \left(\vec e_z\times \nabla_\bot
\frac{\partial \phi}{\partial t} \cdot \nabp\right) \nabp^2 \phi+ \frac{2}{B_0} \left(\vec e_z\times \nabla_\bot \phi
\cdot \nabp\right) \frac{\partial \nabp^2 \phi}{\partial t} \right.
\]
 \be
\left.+ \frac{1}{B_0^2} \left(\vec e_z\times \nabla_\bot  \phi \cdot \nabp\right)  \left[  \left(\vec e_z\times
\nabla_\bot
 \phi \cdot \nabp\right) \nabp^2 \phi \right] \right\}.
\label{e11} \ee
On the right hand side in Eq.~(\ref{e11}), $\phi$ should be taken as  the summation (\ref{s}). In view of the
approximative calculation of the time-varying mode amplitude,  on the right-hand side  the remaining  time and space
derivatives $\partial/\partial t, \nabla_\bot$ imply $\pm  i \omega_l, \pm \vec k_l$ ($l=1, 2, 3$), where $\pm$ appears
due to complex-conjugate expressions, and  out of all terms introduced by the summation (\ref{s}), one should keep only
the resonant ones, corresponding to the $-i\omega_j t + i \vec k_j \cdot \vec r$ on the left-hand side.

Without any loss of generality we further assume the following resonant conditions: \be
\omega_1=\omega_2+ \omega_3, \quad \vec k_1= \vec k_2+ \vec k_3. \label{e9} \ee
The remaining task of calculating the nonlinear terms in Eq.~(\ref{e11})  is presented below. For the mode $\omega_1,
\vec k_1$, from the first term  on the right-hand side in  Eq.~(\ref{e11})  we have
\[
\Gamma_{1,1}\equiv \frac{1}{B_0} \left(\vec e_z\times \nabla_\bot \frac{\partial \phi}{\partial t} \cdot \nabp\right)
\nabp^2 \phi\Rightarrow \frac{i}{B_0} \vec e_z \cdot(\vec k_3\times \vec k_2) (\omega_2 k_3^2 - \omega_3 k_2^2) \Phi_2
\Phi_3.
\]
Note that without the time derivative this nonlinear term would correspond to its counterpart in the electron-ion
plasma \cite{w}.

Similarly, the second nonlinear term yields:
\[
\Gamma_{1,2}\equiv \frac{2}{B_0} \left(\vec e_z\times \nabla_\bot \phi \cdot \nabp\right) \frac{\partial \nabp^2
\phi}{\partial t} \Rightarrow \frac{2i}{B_0} \vec e_z \cdot (\vec k_3\times \vec k_2) (\omega_3 k_3^2 - \omega_2 k_2^2)
\Phi_2 \Phi_3.
\]
Consequently, Eq.~(\ref{e11}) for the mode $\omega_1, \vec k_1$ becomes
\be
\frac{\delta \Phi_1(t)}{\delta t} -  \Upsilon_{13}= -\frac{\vec e_z\cdot (\vec k_3\times\vec k_2)}{2B_0 \omega_1 k_1^2}
\left[\omega_2 k_3^2 -\omega_3 k_2^2+ 2 (\omega_3 k_3^2 - \omega_2 k_2^2)\right] \Phi_2 \Phi_3. \label{e12} \ee
The meaning of the term $\Upsilon_{13}$ is obvious from Eq.~(\ref{e11}), it is the third nonlinear term there.

 In the same manner from the time evolution Eq.~(\ref{e11}) for the modes $\omega_2, \vec k_2$, and $\omega_3, \vec k_3$ one
obtains respectively:
 \be
\frac{\delta \Phi_2(t)}{\delta t} -   \Upsilon_{23}= -\frac{\vec e_z\cdot (\vec k_1\times\vec k_3)}{2B_0 \omega_2
k_2^2} \left[\omega_1 k_3^2 +\omega_3 k_1^2 - 2 (\omega_3 k_3^2 + \omega_1 k_1^2)\right] \Phi_1 \Phi_3^*. \label{e13}
 \ee
\be
\frac{\delta \Phi_3(t)}{\delta t}-  \Upsilon_{33}= -\frac{\vec e_z \cdot (\vec k_1\times\vec k_2)}{2B_0 \omega_3 k_3^2}
\left[\omega_1 k_2^2 +\omega_2 k_1^2 - 2 (\omega_2 k_2^2 + \omega_1 k_1^2)\right] \Phi_1 \Phi_2^*. \label{e14} \ee
Equations~(\ref{e12})-(\ref{e14}) describe time evolution of the three modes due to their nonlinear interaction. For an
arbitrary $j$th mode on the left-hand sides in Eqs.~(\ref{e12})-(\ref{e14}), the first and second nonlinear terms on
the right-hand sides must include terms $k$ and $l$, where $k \neq l$ (including the combination with the
complex-conjugate counterparts too).

On the other hand, the third nonlinear term $ \Upsilon_{j3}$ includes products of all three amplitudes $\Phi_j, \Phi_k,
\Phi_l$, and consequently,  it can not contain resonant exponential terms that would follow from the interaction
between different modes. In other words, for the $j$th mode on the left-hand sides in Eqs.~(\ref{e12})-(\ref{e14}), the
third nonlinear term $ \Upsilon_{j3}$  can only contain the self-interacting terms of the type $c_j\cdot \Phi_j$, where
in the same time the interaction coefficient can  only include terms of the form $\Phi_k \Phi_k^*$, $\Phi_l \Phi_l^*$,
where $k\neq j$, and $l\neq j$. Hence, although it describes the variation of the mode amplitude due to 3-wave
interaction, it does not directly contribute  to a possible energy transfer towards larger and shorter wave-lengths. It
is a cubic nonlinearity yielding only  a  frequency shifts due to the modulational interaction. As such it in principle
introduces a mismatch in the perfect resonant  condition (\ref{e9}) for the frequencies,  yet these effects  will not
be discussed here.

The double energy transfer can follow only from the first and second nonlinear terms in Eq.~(\ref{e11}), therefore only
the contribution of those terms will  be checked below. For that purpose, using the resonant conditions (\ref{e9}) we
rewrite Eqs.~(\ref{e12})-(\ref{e14}) in a more symmetric form
\be
\frac{\delta \Phi_1(t)}{\delta t}-  \Upsilon_{13}= \frac{\vec e_z \cdot (\vec k_2\times\vec k_3)}{2B_0 \omega_1 k_1^2}
\left[\omega_2 k_3^2 -\omega_3 k_2^2+ 2 (\omega_3 k_3^2 - \omega_2 k_2^2)\right] \Phi_2 \Phi_3. \label{e12a} \ee
 \be
\frac{\delta \Phi_2(t)}{\delta t}-  \Upsilon_{23}= \frac{\vec e_z \cdot (\vec k_2\times\vec k_3)}{2B_0 \omega_2 k_2^2}
\left[-\omega_1 k_3^2 -\omega_3 k_1^2 + 2 (\omega_3 k_3^2 + \omega_1 k_1^2) \right] \Phi_1 \Phi_3^*. \label{e13a} \ee
\be
\frac{\delta \Phi_3(t)}{\delta t} -   \Upsilon_{33}= \frac{\vec e_z \cdot (\vec k_2\times\vec k_3)}{2B_0 \omega_3
k_3^2} \left[\omega_1 k_2^2 +\omega_2 k_1^2 - 2 (\omega_2 k_2^2 + \omega_1 k_1^2)\right] \Phi_1 \Phi_2^*. \label{e14a}
\ee
It is interesting to compare these equations  with drift-wave equations in electron-ion plasma \cite{w}, where the
terms $ \Upsilon_{jk}$ are absent and the interaction coefficients for the three modes (without some unimportant common
terms) are given, respectively, by:
\be
\alpha_1=\vec e_z \cdot(\vec k_2\times \vec k_3) (k_3^2-k_2^2) \Phi_2\Phi_3 \label{ei1} \ee
 \be
\alpha_2= \vec e_z \cdot (\vec k_2\times \vec k_3) (k_1^2-k_3^2) \Phi_1\Phi_3^*, \label{ei2} \ee
\be
\alpha_3=\vec e_z \cdot (\vec k_2\times \vec k_3) (k_2^2-k_1^2) \Phi_1\Phi_2^*. \label{ei3} \ee
 The double energy transfer implies an energy  transfer towards both shorter (the direct one) and longer scales (the inverse transfer).
   Taking as an example
\be
k_2^2<k_3^2<k_1^2,\label{k} \ee
for an electron-ion plasma,  from Eqs. (\ref{ei1})-(\ref{ei3}) it turns out that $\alpha_3<0$, implying  the double
energy transfer and the precipitation of energy from the intermediate mode $k_3$ to the other two modes. In the
opposite case $k_2^2>k_3^2>k_1^2$, we have $\alpha_3>0$, $\alpha_{1, 2}<0$ and the intermediate mode receives the
energy from the two others.

Similarly, for
\be
k_3^2>k_2^2>k_1^2,\label{k22} \ee
from Eqs. (\ref{ei1}-\ref{ei3}) it is seen that the intermediate mode $k_2$ loses the energy, $\alpha_2<0$, and the
double energy transfer takes place.

 However, in our case dealing with pair-plasmas, for positive
frequencies and after using (\ref{e9}) to eliminate $\omega_3$, from Eqs.~(\ref{e12a})-(\ref{e14a}) (after disregarding
the obvious positive and common terms) the signs of the coefficients of interaction are determined respectively by:
\be
\beta_1=k_3^2 (2 \omega_1 - \omega_2) - k_2^2 (\omega_1+\omega_2), \label{b1} \ee
\be
\beta_2=k_3^2 ( \omega_1 - 2 \omega_2) + k_1^2 (\omega_1+\omega_2), \label{b2} \ee
\be
\beta_3=k_2^2 ( \omega_1 - 2 \omega_2) + k_1^2 (\omega_2 -2 \omega_1). \label{b3} \ee
As compared with the coefficients $\alpha_j$ from  an e-i plasma, the coefficients  $\beta_j$ appear to be more
complex, with a dependence on frequencies too.

For a more detailed  comparison with the e-i plasma, we take the case (\ref{k}) given above. In order to have the
double energy transfer with the intermediate mode yielding the energy, $\beta_3<0$, $\beta_{1,2}>0$, from Eqs.
(\ref{b1})-(\ref{b3}) we obtain the following additional conditions for the frequencies and wave-numbers: \be
\omega_1>2 \omega_2, \quad k_1^2>k_2^2\, \frac{\omega_1- 2 \omega_2}{2 \omega_1 - \omega_2}, \quad k_3^2>k_2^2\,
\frac{\omega_1+\omega_2}{2\omega_1-\omega_2}. \label{ad1} \ee
Hence, the same double energy transfer in the pair-plasma imposes three more conditions. Yet, it is still possible and
as an example we take the following set of numbers: $\omega_1=3$, $\omega_2=1$, $\omega_3=2$ and $k_1^2=3$, $k_2^2=1$,
$k_3^2=2$. These indeed satisfy all the conditions (\ref{e9}), (\ref{k}), (\ref{ad1}) and allow for the double energy
transfer because: $\beta_1=6$, $\beta_2=14$, $\beta_3=-14$.

Making one more comparison with the e-i plasma, we now use the condition (\ref{k22}) and analyze the coefficients
(\ref{b1})-(\ref{b3}). The requirement $\beta_2<0$, in view of (\ref{e9}) now yields
\be
 2\omega_2>\omega_1>\omega_2,\quad  \mbox{and} \quad k_3^2> k_1^2
(\omega_1+\omega_2)/(2\omega_2-\omega_1). \label{add1} \ee
 The condition $\beta_1>0$ yields $k_3^2>k_2^2(\omega_1+\omega_2)/(2
\omega_1-\omega_2)$, and $\omega_1>\omega_2/2$, the latter being always satisfied in view of the condition (\ref{e9}).
However, the  obtained condition for the frequencies (\ref{add1}) in fact makes the third required  condition
$\beta_3>0$ impossible. Hence, contrary to the e-i plasma case which gives the double energy transfer, in the
pair-plasma  for this case it is absent.

In general, the main reason for these differences is  the presence of frequencies in Eqs. (\ref{b1})-(\ref{b3}),
originating from the necessarily different decoupling (as compared to the e-i plasma) used in the derivations of Eq.
(\ref{e7}).

To conclude, the pair plasma introduces a lot of new interesting physical phenomena, the  nonlinear three-wave
interaction discussed in this Brief Communication being yet another example of it. We stress that pair properties imply
that the results presented here are valid for both homogeneous and inhomogeneous environments.

\begin{acknowledgments}
The  results presented here  are  obtained in the framework of the projects G.0304.07 (FWO-Vlaanderen), C~90347
(Prodex),  GOA/2009-009 (K.U.Leuven). Financial support by the European Commission through the SOLAIRE Network
(MTRN-CT-2006-035484) is gratefully acknowledged.

\end{acknowledgments}

%\subsection{}
%\subsubsection{}

% If in two-column mode, this environment will change to single-column format so that long equations can be displayed.
%\begin{widetext}
%$$\mbox{put long equation here}$$
%\end{widetext}

% Figures should be put into the text as floats.
% Use the graphics or graphicx packages (distributed with LaTeX2e).
% Put the label that you will use with \ref{} command in the braces of the \label{} command.
%
% \begin{figure}
% \end{figure}

% \begin{table}
% \end{tabular}
% \end{table}

\vfill

\pagebreak

\end{document}